\documentstyle[epsfig1,emulateapj]{article}

\newcommand{\etal}{{\it et al.\ }}

\newcommand{\avg}[1]{\langle{#1}\rangle}

\newcommand{\ltsima}{$\; \buildrel < \over \sim \;$}
\newcommand{\lsim}{\lower.5ex\hbox{\ltsima}}
\newcommand{\gtsima}{$\; \buildrel > \over \sim \;$}
\newcommand{\gsim}{\lower.5ex\hbox{\gtsima}}

\begin{document}


\title{Fast CMB Analyses via Correlation Functions}

\author{Istv\'an Szapudi$^1$, Simon Prunet$^1$, Dmitry Pogosyan$^1$}
\author{Alexander S. Szalay$^2$, and J. Richard Bond$^1$}
\affil{1. CITA, University of Toronto, 60 St George St,
Toronto, Ontario, M5S 3H8, Canada}
\affil{2. Department of Physics and Astronomy, Johns Hopkins University,
Baltimore, MD 21218, USA}



\begin{abstract}


We propose and implement a fast, universally applicable method 
for extracting the angular power spectrum ${\cal C}_{\ell}$ 
from CMB temperature maps by first estimating the 
correlation function $ \xi(\theta)$. Our procedure recovers
the ${\cal C}_{\ell}$'s 
using $N^2$ (but potentially $N \log N$), operations,
where $N$ is the number of pixels.
This is in contrast with standard maximum likelihood techniques 
which require $N^3$ operations. Our method makes no special 
assumptions about the map, unlike
present fast techniques which rely on symmetries of the underlying
noise  matrix, sky coverage, scanning strategy, and geometry.
This enables for the first time the analysis
of megapixel maps without symmetries.
The key element of our technique is
the accurate multipole decomposition of $\xi (\theta)$.  The ${\cal
C}_{\ell}$ error bars and cross-correlations are found by a Monte-Carlo
approach.  We applied our technique to a large number of
simulated maps with Boomerang sky coverage in $81000$ pixels.
We used a diagonal noise matrix, with approximately
the same amplitude as Boomerang. These studies demonstrate that our
technique provides an unbiased estimator of the ${\cal C}_{\ell}$'s. Even
though our method is approximate, the error bars obtained are nearly optimal,
and converged only after few tens of Monte-Carlo realizations. 
Our method is directly applicable for the non-diagonal noise matrix. This, 
and other generalizations, such as minimum variance weighting
schemes, polarization, and higher order
statistics are also discussed.

\end{abstract}





\section{Introduction}

Future missions of measuring the Cosmic Microwave Background (CMB)
fluctuations will revolutionize our knowledge of cosmology. They
will either confirm or refute the basic Big Bang paradigm, 
will reveal the values of most cosmological parameters within
a few percent (e.g., Spergel 1994, Knox 1995, Hinshaw, Bennett, \& Kogut
1995, Jungman \etal 1996, Zaldarriaga, Spergel, \& Seljak 1997, Bond, 
Efstathiou, \& Tegmark 1997).
The large number of pixels contained in current and future experiments
enable these exciting developments, but at the same time, 
present unprecedented challenges the data analysis.
The mainstream maximum likelihood techniques are already pushed
to their limits by the largest existing CMB maps, but they
are clearly inadequate for future megapixel surveys.
Therefore the most important near term task
for CMB research is to find techniques which could
perform the required analyses with realistic resources, thus
fulfill the promise of the high precision experiments.

This Letter proposes a fast, universally applicable
method for the estimation
of $C_l$'s from any CMB pixel map, based on estimating
correlation functions. The recovered errorbars are
at most $10\%$ larger than the theoretically smallest possible ones
imposed by cosmic variance and noise. Optimal methods could further
decrease these errorbars only slightly, and at an unrealistic cost; thus
they  represent a case of diminishing returns.
Our procedure scales as $N^2$, 
and more sophisticated algorithms will improve this to $N \log N$. 
This essentially solves the problem for two major
upcoming space missions, MAP (Microwave
Anisotropy Probe) and the Planck, and
at the same time allows more detailed analyses
of current and future balloon-borne and ground based experiments.
The short analysis turn around time  enables
Monte-Carlo (MC) studies of systematics, foregrounds, errors, underlying models
etc.  These are essential to the interpretation of the data but
inaccessible to present methods.

The standard maximum likelihood
methods for analyses were developed and tested on COBE measurements
(e.g., G\'orski 1994, G\'orski \etal 1994, 1996, Bond 1995, Tegmark
\& Bunn 1995, Hinshaw \etal 1996, Tegmark 1996, Bunn \& White 1996,
Bond, Jaffe, \& Knox 1998, hereafter BJK98, 2000) 
which have only about $N \simeq 1000$ pixels. Balloon-borne experiments, 
such as Boomerang (de Bernardis \etal 2000), Maxima (Hanany \etal 2000),  and 
Tophat (Martin \etal 1996),  and ground based measurements e.g., 
TOCO (Miller \etal 1999) and Viper (Peterson \etal 2000)
with up to $N \simeq 10^5$ are already
pushing present day supercomputer technology. The future
missions MAP and Planck with $N = 10^6-10^7$ are estimated to require
up to millions of years 
(Borrill 1999, Bond, Crittenden, Jaffe, \& Knox 2000)
for one iteration of the quadratic maximum likelihood estimator
for $C_l$'s. The disk storage and
memory usage of these algorithms are prohibitive as well.

To date fast algorithms were presented under 
two fairly restrictive assumptions:
i) the noise is both temporally uncorrelated and spatially axially symmetric 
ii) the foregrounds can be exactly removed both from the map 
and the correlation matrix
(Oh, Spergel, \& Hinshaw 1999, Wandelt, Hivon \& Gorski 1998, 2000).
Our approach is significantly different from these, since it
does not assume any symmetries.



\section{Description of the Method}
\begin{figure*} [htb]
\centerline{\hbox{\psfig{figure=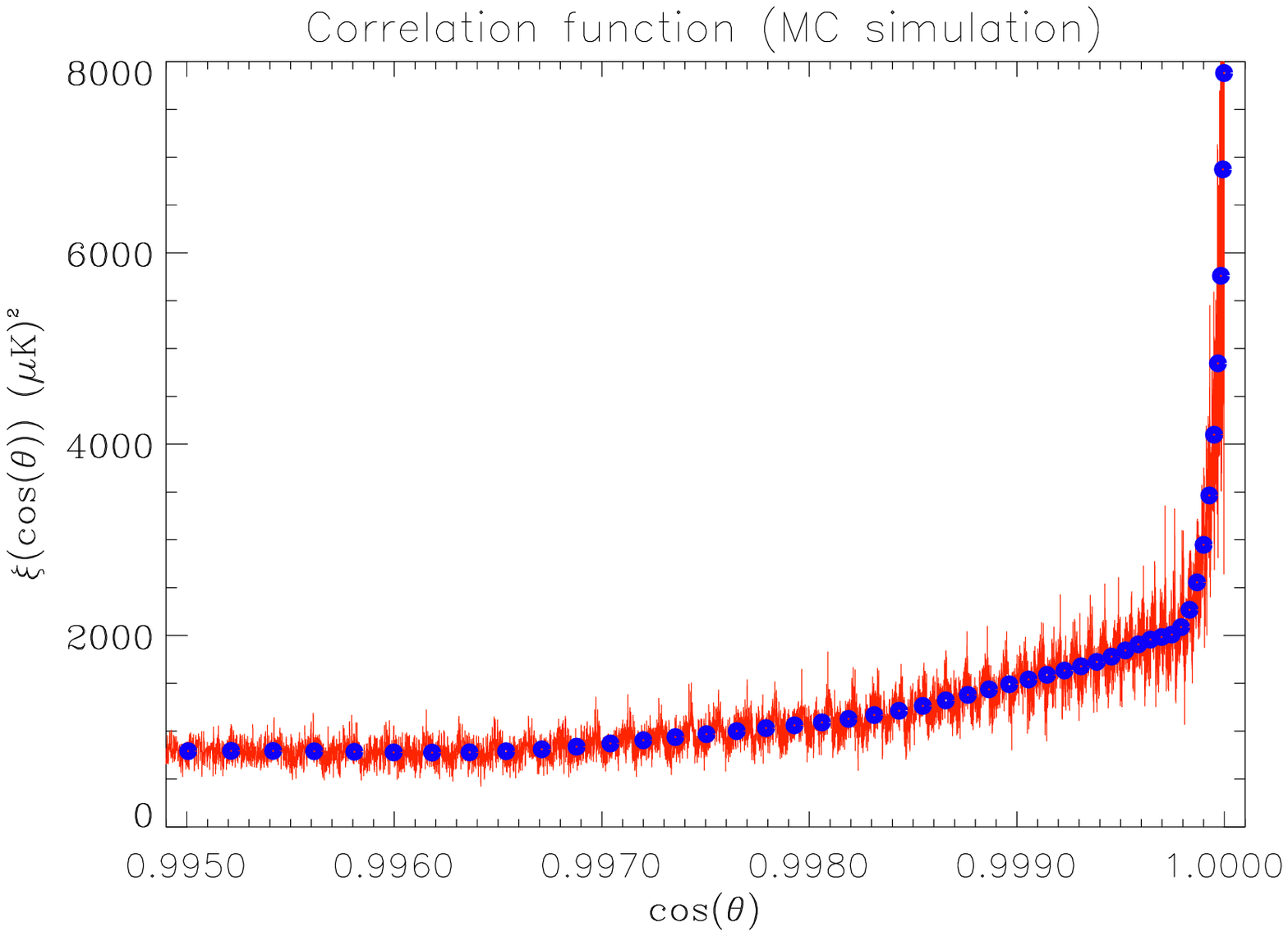,width=9cm},
\psfig{figure=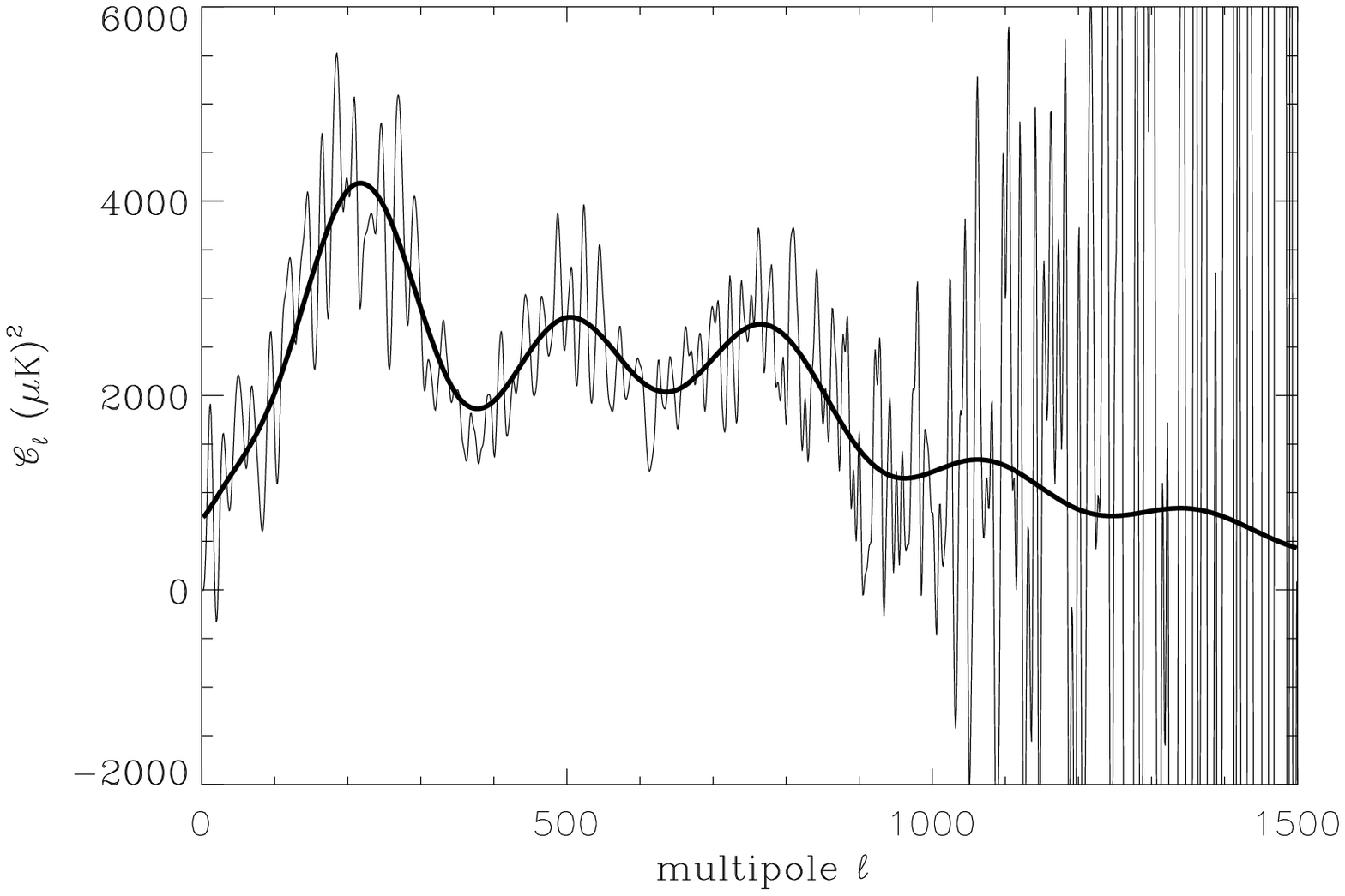,width=9cm} }}
\caption{The left panel displays a typical high resolution
measurement of the two-point correlation function along
with its Gaussian resampling at Legendre roots (large dots).
The singular feature at around $1$ degree ($\cos\theta\simeq 0.9998$)
condenses most of the information about the $C_l$'s. The right
panel present a few examples of the inversion process.
Individual $C_l$'s are calculated from the correlation function
with the method described in the paper.
}
\label{fig1}
\end{figure*}       

The basic steps of our recipe to extract $C_l$'s from
a large CMB map are conceptually simple: first we measure
the two-point correlation function in high resolution
bins, next we smooth it with Gaussian kernels centered
on the roots of Legendre polynomials, finally, we integrate
to obtain the $C_l$'s. Each step is fine-tuned in order to
arrive at a fast method, which is as precise as possible.

Let us denote the temperature fluctuations at a sky vector
$q$, a unit vector pointing to a pixel on the sky, with
$T(q)$. In isotropic universes the two-point correlation
function is a function only of the angle between the
two vectors and can expanded into a Legendre series,
\begin{equation}
  \xi_{12} = \avg{T(q_1)T(q_2)} = 
  \sum_\ell \frac{\ell + 1/2}{\ell ( \ell+1)} {\cal C}_\ell P_\ell(\cos\theta),
  \label{eq:xi}
\end{equation}
where $\cos\theta = q_1 . q_2$, the dot product of the two unit vectors,
$P_\ell(x)$ is the $\ell$-th Legendre polynomial, and the 
${\cal C}_\ell$ coefficients
realize the angular power spectrum of fluctuations. If the 
CMB anisotropy is Gaussian, which is expected to be
an excellent approximation, the correlation function,
or the ${\cal C}_\ell$'s yield full statistical description.

In reality each pixel value, $\Delta_i$, contains contributions
from the CMB and noise; the latter is also assumed to be
Gaussian with a correlation matrix (the noise matrix) $N_{ij}$,
determined during map-making. The full pixel-pixel correlation
matrix is
  $C_{ij} = \xi_{ij}+N_{ij}.$
We adopted the edge and noise corrected
estimator of Szapudi \& Szalay (1998),
\begin{equation}
  \tilde\xi(\cos\theta) = \sum_{ij} f_{ij} (\Delta_i \Delta_j - N_{ij}),
  \label{eq:estimator}
\end{equation}
where $f_{ij} = 0$ unless the pair of pixels belong to a particular
bin in $\cos\theta$, and $\sum_{ij}f_{ij} = 1$.
The above estimator is unbiased, i.e. $\avg{\tilde\xi(\cos\theta)}
= \xi(\cos\theta)$. A complex pair weighting might be required
in surveys with uneven coverage, and to optimize the performance
of the estimator (see discussions in \S 4).
For this Letter we used a straightforward $N^2$
program to estimate the above quantity in a large number of
bins linear in $\theta$ massively oversampling the
pixel separation. The results are robust against
the exact number of bins used, for the measurements presented here
we have placed $300,000$ bins in the range of $0-\pi$.

The raw correlation function is then resampled
with a Gaussian filter with additional weighting proportional to the
number of pairs in each bin. 
The centers of the filters are determined
by the roots of the Legendre polynomial $P_{\ell_{max}}$ of the
highest $\ell_{max}$ to be measured; their widths is about half
the pixel size. This procedure allows accurate centering of the
filters, suppresses any high frequency 
sub-beam measurement noise due to the massive oversampling. 
The Gaussian convolution
translates into a simple multiplication in $\ell$-space, thus easy
to correct for. 
We verified that, after this correction,
the final result is robust against varying the
size of the filter; we have used $4'$. 
For small surveys, a smooth cut-off
of the correlation function is necessary on large angles
to suppress
noise from edge effects. We checked that the ${\cal C}_\ell$'s are 
robust against changing the scale and sharpness of the cut-off
(up to a corresponding smoothing in $\ell$-space);
we have used exponential tapering above $18^\circ$.
The left panel of Figure~\ref{fig1} shows an example of a
measured correlation functions along with its resampled version
(large dots). The ${\cal C}_\ell$'s are determined from the resampled 
correlation function by Legendre-Gauss integration (e.g., Press \etal 1992).
Since this is exact up to $2 \ell_{max} +1$, 
the ${\cal C}_\ell$'s are as accurately
recovered as the correlation function.
The right panel of Figure~\ref{fig1} shows examples of ${\cal C}_\ell$'s
inverted
from two-point correlation function measurements in simulations.

The full correlation matrix of the above estimator can be 
calculated analytically
as well. Under Gaussian assumption, the cross-correlation
between two bins denoted with $1$ and $2$, respectively is
  $\avg{\delta\xi_1\delta\xi_2}= 
   \sum_{ijkl} f^1_{ij}f^2_{kl}[\xi_{ik}(\xi_{jl}+N_{jl})+k\leftrightarrow l)].
$
The straightforward calculation of the above expression 
takes $N^4$ operations, thus infeasible for a megapixel survey.
While this scaling is likely to be improved to  $N(\log N)^3$
with the advanced algorithms in the near future, 
in this Letter we propose
an entirely different approach: high volume MC simulations.
Certain systematics
can only be taken into account this way, and the speed of our method
makes such computations possible.

\section{Simulations}

\begin{figure*} [htb]
\centerline{\hbox{\epsfig{figure=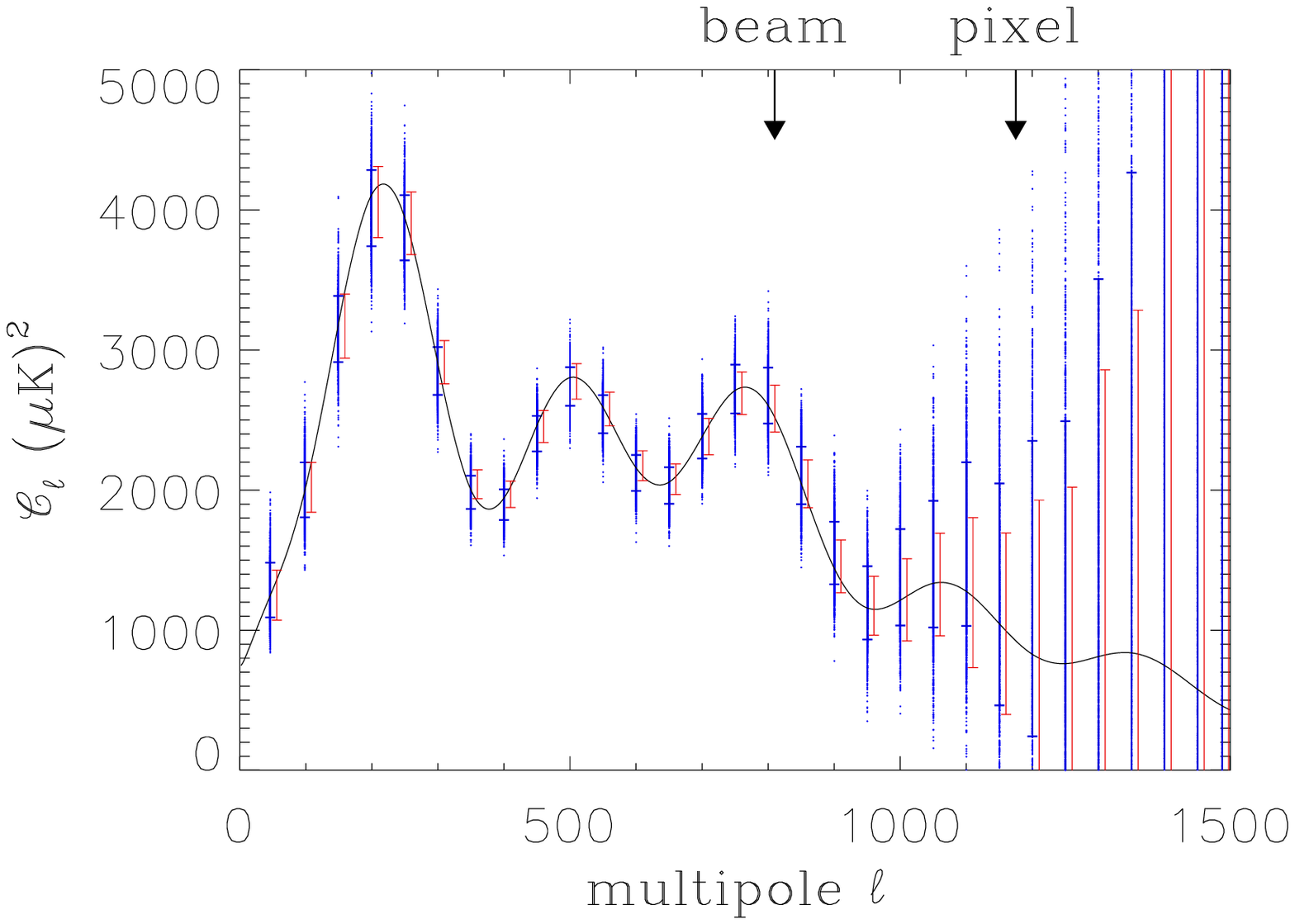,width=9cm},
\epsfig{figure=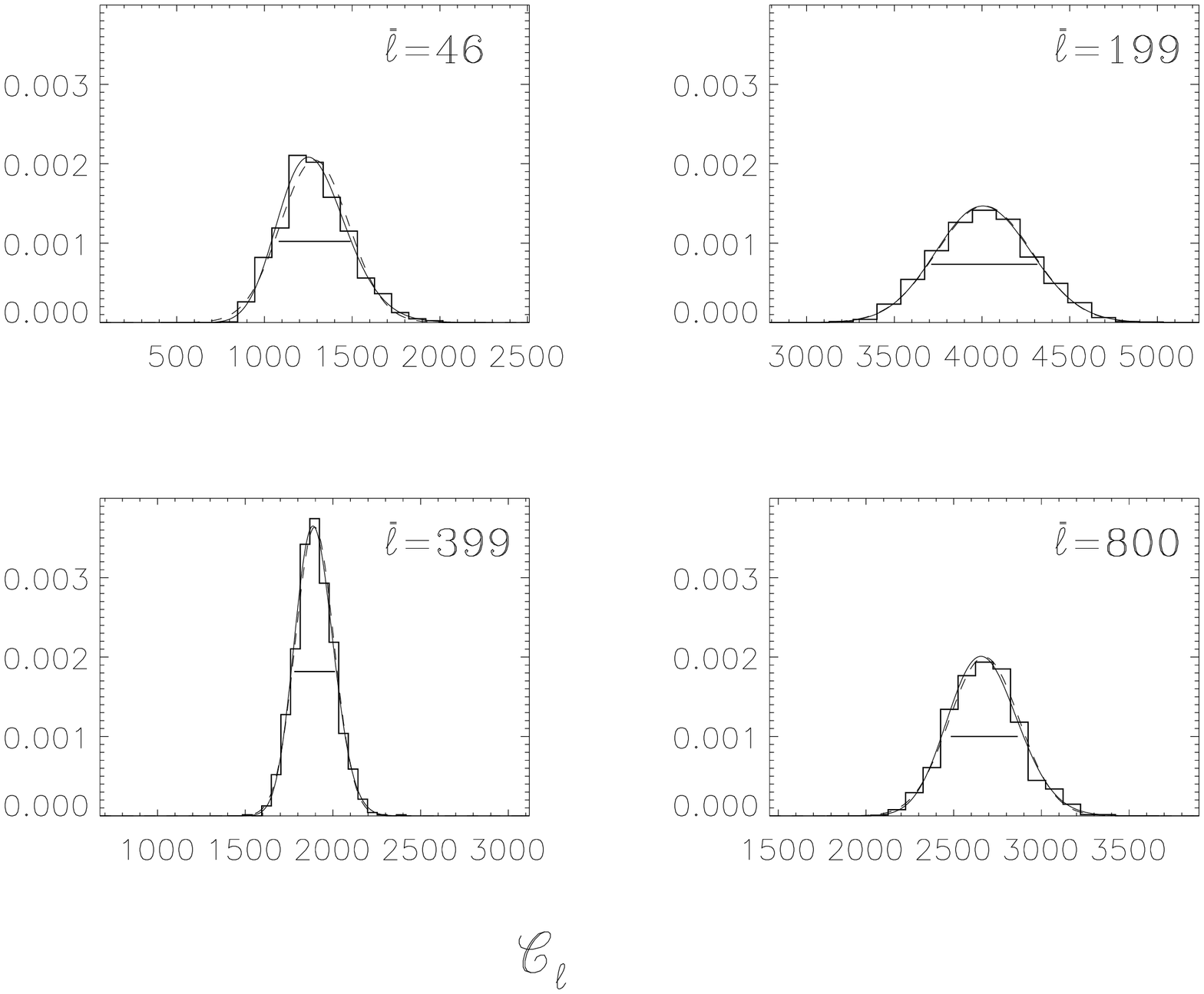,width=5.625cm,angle=90} }}
\caption{ The left panel displays the main results of this
Letter: ${\cal C}_\ell$'s calculated in 1298 Boomerang-like simulations
and then rebinned into flat ${\cal C}_\ell$-bands with width of $50$.
The small points show the individual measurements with
errorbars representing the standard deviations in each band. 
Theoretical errorbars of Equation~\ref{eq:errorbar} are displayed
shifted to the right for  clarity. The comparison shows that our 
method is unbiased, and has nearly optimal variance. The arrows
point to the effective beam and pixel scales.
The right
panel shows the distribution of measurements in selected bands.
Shifted lognormal (dash) and Gaussian curves are fitted to the
histogram. Except for the $\ell=50$ band, the two are indistinguishable. 
Horizontal lines display optimal FWHM dispersion for comparison.
}
\label{fig2}
\end{figure*}       
%
%


To illustrate the method, we generated $1298$ MC realizations
of CMB maps with a standard CDM power spectrum and typical 
instrumental noise.
The sky coverage used to create the
maps ($\sim 1000 \,{\rm deg}^2$), the noise level 
($\sim45 \,\mu{\rm K}$ per pixel), 
and the Gaussian beam size ($10'$ FWHM) were comparable to the BOOMERanG 
experiment (de Bernardis \etal 2000). Full sky maps 
in $\sim 7'$ pixels were
generated  by the  HEALPix (Gorski \etal 1998) {\tt synfast} routine.
Then maps were reduced to the desired coverage, and 
uniform white noise has been added.

The angular power spectra have then been extracted, corrected for 
beam, pixelization, and the additional $4'$ smoothing of the
correlation function,  and finally averaged into flat band-powers 
${\cal C}_{\bar \ell} $ with widths $\Delta \ell = 50$ and
positions $\bar \ell$ that of de Bernardis \etal (2000).
Figure~\ref{fig2} shows ${\cal C}_{\bar \ell}$ for all realizations,
together with the corresponding error bars. They are to be compared with
the optimal error bars computed from the approximate formula
\begin{equation} 
\Delta{\cal C}_\ell = \sqrt{\frac{2}{(2\ell+1)f_{sky}}}\left(
{\cal C}_\ell + w^{-1}\ell(\ell+1)W^{-2}_\ell/2\pi \right),
\label{eq:errorbar}
\end{equation}
where $f_{sky}$ is fraction of the sky observed, $w^{-1}$ is the noise variance
and $W_{\ell}$ is the beam of the experiment.
According to Figure~\ref{fig2}, our estimator is nearly optimal and 
unbiased. In any given $\bar \ell$-band, 
the differences between the measured and theoretical
mean and variance are at most $10\%$ of the optimal variance.
The MC method for determining the errorbars converges
remarkably fast: a few tens of realizations give an accurate estimate
of both the mean and variance. For real data, the smoothed
measured ${\cal C}_\ell$'s themselves should be taken for 
generating MC realizations.

In addition,  we have estimated the cross-correlations between bands,
and found them to be consistent with zero, decreasing as
 $\simeq 3/\sqrt{N_r}$ with the number of realizations $N_r$.
These experiments show that our 
method does not induce spurious correlations into the measurements.

The right panel of Figure~\ref{fig2} displays distributions 
of several recovered 
band-powers, illustrating different regimes of
signal or noise domination of the total variance.
We found that the distribution in all bands is well fitted with
an offset lognormal probability density (BJK98), although it is 
indistinguishable from a Gaussian in most bands.

In summary, our method has nearly identical mean, variance,
cross-correlations, and distribution as one would expect from
standard maximum likelihood methods. The present implementation
of the above pipeline takes about 25 minutes of serial CPU
on a small workstation, including artificial map generation,
measurement of the correlation function, and estimating
the $C_l$'s individually, and finally summarizing them into band-powers.
The required time is dominated by the measurement of the two-point
correlation function (about 20 minutes); obtaining the $C_l$'s
from it is negligible (few seconds).

These measurements illustrate that pairwise estimators with heuristic
weights are feasible, their speed is far superior
to maximum likelihood techniques, and they can be rendered nearly optimal.
While details of the above recipe are subject to honing for
future practical applications, our 
present code is capable of analyzing MAP 
on a sub-supercomputer class CPU without explicit reliance on symmetries;  
a feat no other method could claim.

\section{Discussion}

We presented a novel method to extract  ${\cal C}_\ell$'s from 
large CMB maps via two-point correlation functions.  
We have cast our determination of ${\cal C}_\ell$ into 3 steps:
fine-grained $C(\theta)$ estimation, the pixel-related $4^\prime$
Gaussian smoothing of it, followed by $P_\ell$ multiplication and
Gauss-Legendre integration. The resulting expression for ${\cal
C}_\ell$ can be itself understood as  a sum over pixel pairs, with a
pair-weighting proportional to the Gaussian smoothing of $P_\ell$.
In this respect our estimator belongs to the class of quadratic estimators
for ${\cal C}_\ell$, of which other examples (e.g.,
BJK98) have been used in the
past and are expected to give similar results.


In the present example the measured errorbars
were at most $10$\% larger then the theoretically minimal
ones; this negligible suboptimality allowed the 
reduction of the CPU time from months (Borrill 1999, Table 3)
to hours. For future large missions the 
difference will be even more dramatic: optimal methods
would improve the errorbars only slightly at a prohibitive
cost. 

Our new approach has several advantages in comparison 
to previous techniques.
It scales as $N^2$ even in its most straightforward
implementation, compared to typical quadratic estimators which
scale as $N^3$. This scaling can be further improved up to
$N\log N$.
While in the present implementation we used diagonal noise,
no other (e.g., asymuthal) symmetries were assumed about noise,
or geometry of the map. Cut out holes around bright sources,
galactic cut, any irregularity in the sampling,
make no difference in the speed or performance of the algorithm.
To illustrate this, Equation~\ref{eq:estimator} 
was used to determine the correlation function from COBE DMR data,
which has inhomogeneous noise. The recovered $C_l$'s are consistent with those
obtained from implementations of  quadratic estimators discussed in BJK98.
In contrast with any other previous attempt, our
approach is straightforward to generalize for
non-diagonal noise matrix (see below).

While our method in its present form is already practical for
analyzing megapixel CMB maps, it has  the potential for further
generalizations.
General non-diagonal noise appears to pose a serious problem. 
Fast iterative map making methods (Wright \etal 1996,
Prunet \etal 2000)  capable of handling large data sets 
only furnish the weight matrix, $w_{ij} = N_{ij}^{-1}$.
Since inversion of the noise matrix is again an $N^3$ problem,
our estimator of Equation~\ref{eq:estimator} might not be
directly applicable. Instead, we propose a new estimator 
using MC realizations of artificial noise 
\begin{equation}
  \tilde\xi(\cos\theta) = \sum_{ij} f_{ij} (\Delta_i \Delta_j - 
    \frac{1}{M}\sum_{k=1}^M n_i^k n_j^k),
  \label{eq:modestimator}
\end{equation}
where $n_i^k$ is one of $M$ realizations of the noise for pixel $k$. 
This plays the role of a
MC inversion of the weight matrix $w_{ij} = N_{ij}^{-1}$.
Generation of $n_i^k$ in the time domain, where it has simple
correlation structure, is straightforward. Its iterative projection
into pixel space is equivalent to
map-making. This is feasible even when storing
the noise-matrix would be prohibitive, as for
Planck.

To further improve the performance of our method, we will
implement a  minimum variance
weighting scheme  (Feldman Kaiser, \& Peacock 1994,
and Colombi, Szapudi, \& Szalay 1998). This corresponds to down-weighting
measurements with their variance, and
might be important in maps, where various
pairs contributing to the correlation function have widely differing errors.
Otherwise, the uniform weighting scheme is nearly optimal 
(Colombi, Szapudi, \& Szalay 1998). The present simple weighting 
scheme can be improved heuristically via
individual pixel weighting reflecting differences in sky coverage. 
More complex pair weighting can be defined iteratively,
using the MC estimates of the variances. 

The present Letter used a simple $N^2$ code to calculate
the correlation function, and this is perfectly adequate for LDB
surveys with $N\simeq 10^5$ pixels, and, with supercomputers, even
for MAP. Nevertheless, the development of an $N\log N$ code is
under way (Szapudi \& Colombi 2000, Connolly, Nichol, Moore, \& Szapudi
2000). 

Our technique has reestablished the utility of correlation
functions for CMB studies. This approach has further 
potential applications:
it is naturally generalizable for the assessment
of non-Gaussianity in CMB maps via $k$-point correlation functions,
with implementations as fast as $N(\log N)^{k-1}$
(Sunyaev-Zeldovich effect, lensing of CMB);
it is equally useful for polarization
correlation functions and for obtaining the appropriate $C_l$'s
from them. We have found that the statistical information is 
condensed into singular features of the correlation function,
(see also Bashlinsky \& Bertschinger, in prep.). This
suggests that direct parameter estimation from the CMB two-point correlation
function might be fruitful as well.



Other applications include
correlations of the infrared (SCUBA, BLAST, SIRTIF, FIRST)
and optical background, and weak gravitational lensing.


\begin{thebibliography}{}

\bibitem[]{} Bond J.R., 1995, \prl, 74, 4369 
\bibitem[]{} Bond, J.R., Efstathiou, G., \& Tegmark, M. 1997, \mnras, 291, L33 
\bibitem[]{} Bond, J.R., Jaffe, A.H. \& Knox, L. 1998, \prd, 57, 2117 
\bibitem[]{} Bond, J.R., Jaffe, A.H. \& Knox, L. 2000, \apj, in press
\bibitem[]{} Bond, J.R., Crittenden, G.C., Jaffe, A.H. \& Knox, L. 2000, 
\apj, in press
\bibitem[]{} Borrill, J. 1999, in Proc. 3k Cosmology EC-TMR Conf. 
(eds Langlois, D., Ansari, R., \& Vittorio, N.), 277, (American Institute
of Physics Conf. Proc. Vol 476, Woodbury, New York) 
\bibitem[]{} Bunn, E.F, \& White, M. 1997, \apj, 480, 6 
\bibitem[]{} Colombi S., Szapudi I., Szalay A.S., 1998, MNRAS, 296, 253
\bibitem[]{} Connolly A.J., Moore, A., Nichol, R.C., Szapudi, I., 2000, 
in prep.
\bibitem[]{} de Bernardis, \etal 2000, Nature, 404, 955 
\bibitem[]{} Feldman, H.A., Kaiser, N., \& Peacock, J.A. 1994, \apj, 426, 23 
\bibitem[]{} G\'orski, K.M. 1994, \apj, 430, L85
\bibitem[]{} G\'orski, K.M \etal \& 1994, \apj, 430, L89 
\bibitem[]{} G\'orski, K.M \etal \& 1996, \apj, 464, L11 
\bibitem[]{} G\'orski, K.M, Hivon, E. \& Wandelt, B.D. 1998 
in Proc. MPA/ESO Conf. (eds. Banday, A.J., 
Sheth, R.K., \& Da Costa, L.) (ESO, Garching)
\bibitem[]{} Hanany, S. \etal, (2000) \apjl, submitted (astro-ph/0005123)
\bibitem[]{} Hinshaw, G., Bennett, C.L., \& Kogut, A. 1995, \apj, 441, L1 
\bibitem[]{} Jungman, G., Kaminonkowski, M., Kosowski, A., \& Spergel, D.N. 
1996, \prd, 54, 1332 
Szapudi, I., Szalay, A.S. 2000, \apjl, in press
\bibitem[]{} Knox 1995, \prd, 52, 4307 
         1993, \apj, 412, 64 
\bibitem[]{} Martin, N., \etal 1996, in Space Telescopes and Instruments IV 
Proc.  SPIE (eds. P. Y. Bely and J. B. Breckinridge), 2807, 86 
\bibitem[]{} Miller, A.D., \etal 1999, \apj, 524, 1 
\bibitem[]{} Oh, S.P., Spergel, D.N., \& Hinshaw, G. 1999, \apj, 510, 551 
\bibitem[]{} Peterson, J.B., \etal 2000, \apj, 532, 83 
\bibitem[]{} Press,W.H., Teukolsky,S.A., Vetterling,V.T.  
\& Flannary,B.P. 1992, Numerical Recipes in C,
(Cambridge: Cambridge University Press)
\bibitem[]{} Prunet \etal 2000, in prep 
\bibitem[]{} Tegmark, M. \& Bunn, E.F. 1996, \apj, 464, L35 
\bibitem[]{} Spergel, D.N. 1994, Warner Prize Lecture, BAAS, 185.7301 
\bibitem[Szapudi \& Szalay 1998]{ss98}  Szapudi, I. \& Szalay, A.S. 1998, 
           \apj, 494, L41 (SS)
\bibitem[]{} Szapudi I., Colombi S. 2000, in prep.
\bibitem[]{} Zaldarriaga, M., Spergel, D.N., \& Seljak, U. 1997, \apj, 488, 1 
\bibitem[]{} Wandelt, B.D., Hivon, E. \& G\'orski, K.M 1998,
(astro-ph/9808292)
\bibitem[]{} Wandelt, B.D., Hivon, E. \& G\'orski, K.M 2000, preprint,
(astro-ph/0008111)
\bibitem[]{} Wright, E.L., Hinshaw, G. \& Bennett, C.L. 1996, \apj, 458, L53 

\end{thebibliography}
\end{document}